\documentclass[%
 reprint,
superscriptaddress,
 amsmath,amssymb,
 aps,
]{revtex4-1}

\usepackage{graphicx}
\usepackage{subcaption}
\usepackage{dcolumn}
\usepackage{bm}
\usepackage{upgreek}
\usepackage{epstopdf}

\usepackage{xcolor}

\begin{document}


\title{Johann-type laboratory-scale X-ray absorption spectrometer with versatile detection modes}

\author{Ari-Pekka Honkanen}
 \email{ari-pekka.honkanen@helsinki.fi}
\author{Sami Ollikkala}
\affiliation{%
 Department of Physics, University of Helsinki\\ 
 PO Box 64, FI--00014 Helsinki, Finland
}%
\author{Taru Ahopelto}
\affiliation{%
 Department of Physics, University of Aberdeen\\ 
King's College, Aberdeen, AB24 3FX, United Kingdom
}%
\affiliation{%
 Department of Physics, University of Helsinki\\ 
 PO Box 64, FI--00014 Helsinki, Finland
}%
\author{Antti-Jussi Kallio}
\author{Merja Blomberg}
\author{Simo Huotari}%
\affiliation{%
 Department of Physics, University of Helsinki\\ 
 PO Box 64, FI--00014 Helsinki, Finland
}%




\date{\today}

\begin{abstract}
We present a low-cost laboratory X-ray absorption spectrometer that uses a conventional X-ray tube source and bent Johann-type crystal monochromators. The instrument is designed for XAS studies in the 4--20 keV range which covers most K edges of \textit{3d} transition metals and L edges of \textit{5d} transition metals and actinides. The energy resolution is typically in the range of 1--5 eV at 10~keV depending on the crystal analyser and the Bragg angle. 
Measurements can be performed in transmission, fluorescence, and imaging modes. Due to its simple and modular design, the spectrometer 
can be modified to accommodate additional equipment and complex sample environments
required for \textit{e.g.} \textit{in situ} studies. A showcase of various applications is presented.
\end{abstract}

\maketitle


\section{Introduction\label{sec:introduction}}

The history of laboratory-based X-ray absorption spectroscopy (XAS) instruments
dates back to the discovery of X-rays \cite[pp. 76--78,97]{azaroff_book}. 
The instrument development was rapid in the 1970's \cite{knapp78,bahgat79} with the rise of the extended X-ray fine-structure (EXAFS) technique \cite{sayers71},  but owing to the advent of dedicated synchrotron light sources and the subsequent proliferation of high brilliance facilities \cite[pp. 5--7]{winick95}, the interest toward laboratory instrumentation somewhat faded. Thereafter, XAS has become a well-established technique for obtaining element-specific information on the local structure and chemical state. The use of X-ray spectroscopy in the field of materials science is growing rapidly which has led to a tremendous increase in the demand for synchrotron beamtime during the past decades. As a consequence, one finds it difficult to access synchrotrons for routine measurements which 
could be crucial for understanding the chemistry and structure of novel materials.

In addition to heavy competition, the use of synchrotrons poses other complications as well. Especially with complex \textit{in situ} setups and radioactive materials, transporting the relevant equipment and samples to the beamline and back can be a daunting exercise in logistics and bureaucracy. Managing through all this extra-scientific effort, the experimentalists ought to have a strong desire to ensure that also the scientific prerequisites are \emph{par excellence} before arriving at the beamline. A strong cavalcade of pre-characterisation techniques can thus make a difference between a success and a farce.

For the aforementioned reasons, laboratory-based spectrometers \cite{knapp78,cohen80,georgopoulos81,williams83,thulke83,tohji83,yuryev07,szlachetko13,seidler14,schlesiger15,seidler16,nemeth16} are an interesting option for many applications and they have advanced greatly in energy resolution and throughput. Especially the manufacturing techniques of bent crystal optics have seen major improvements in recent years \cite{bergmann98,collart05,verbeni05, yavas07, verbeni09,rovezzi17,saint-gobain,xrs_tech}. 

In this manuscript we introduce a low-cost laboratory X-ray absorption spectrometer utilising a conventional X-ray tube source and bent Johann-type crystal monochromators. The instrument is designed for XANES and EXAFS studies in the 4--20 keV range which covers most K edges of \textit{3d} transition metals and L edges of \textit{5d} transition metals and actinides. The energy resolution is in the range of 1--5 eV at 10 keV which depends on the analyser crystal and the Bragg angle. Measurements can be performed in the transmission and fluorescence detection modes, the latter enabling the study of thick samples, or samples on a substrate/support. The instrument can be also used for imaging purposes of which a demonstration is given. Due to its simple and modular design, the spectrometer can be modified to accommodate additional equipment and complex sample environments required for \textit{e.g.} \textit{in situ} studies. A showcase of various applications is presented.

\vspace*{-1.735cm}

\section{Instrument design\label{sec:design}}

\begin{figure*}
\centering
\includegraphics[width=\textwidth]{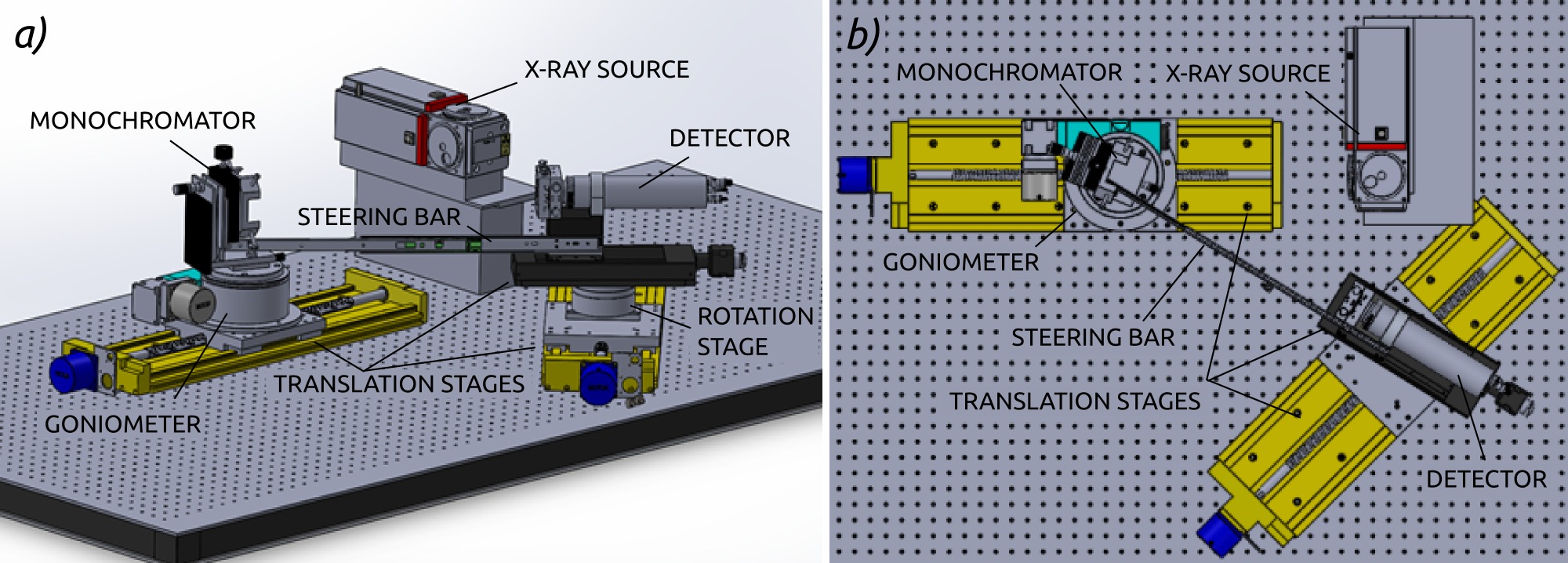}
\caption{A schematic drawing of the essential parts of the instrument from \emph{a)} a perspective view and \emph{b)} above. The helium chamber is not shown for clarity. \label{fig:instrument}}
\end{figure*}

The design is based on the Johann geometry \cite{johann31} in which a polychromatic source of X-rays, a spherically bent crystal analyser and a detector follow the Rowland circle for proper monochromatisation and focusing of X-rays. The diameter of the Rowland circle of the instrument is 0.5 m which is dictated by the bending radius of the analyser crystals. The schematic drawing of the main components is shown in Figure~\ref{fig:instrument}. 

\vspace*{1.735cm}

\subsection{Components}

\subsubsection{Source} 

The X-ray source is a conventional 1.5~kW fine focus X-ray diffraction glass tube. A variety of anode materials can be used. In our case, a typical choice is Ag since it does not have characteristic x-ray lines in the region 3--22~keV and is thus suitable for the operation with the Bremsstrahlung region in the energy range of the design target 4--20~keV. This makes Ag a well suited choice for \textit{3d} transition metal K edge and \textit{5d} L edge spectroscopy. 

The focal spot of the tube is $0.4 \times 8$~mm$^2$ which at 6$^\circ$ take-off angle transforms to the point source size of $0.4 \times 0.8$~mm$^2$, the dimension being perpendicular and parallel to the diffraction plane. The apex angle (or the opening angle) of the X-ray cone is around 7$^\circ$ which covers a circular area of 60~mm in diameter on the monochromator crystal.

\subsubsection{Monochromator and energy resolution} 

For monochromatisation and focusing the polychromatic X-rays 
we use spherically bent crystal analysers (SBCAs) 
provided by ESRF Crystal Analyser Laboratory \citep{rovezzi17}. 
In our case the SBCAs have a bending radius of 0.5~m and a surface diameter of 100~mm. Spherical bending causes strain fields in the monochromator which affect its energy resolution \cite{takagi62,taupin64,takagi69,honkanen14}. To mitigate the effect, the crystal wafers are cut into 15~mm wide strips before they are bonded to the glass concave.
 
In the designed operational range, the typical energy resolution of the analyser is $\gtrsim$ 1~eV \emph{i.e.} $\gtrsim 10^{-4}$ relative to the mean photon energy. Additional contributions of the source size and the other geometric factors, such as the Johann error, to the energy resolution of the instrument were estimated with ray tracing simulations. As shown in Table~\ref{tbl:rel_e_res}, the source size becomes the dominant factor to the energy resolution at the lower Bragg angles. Since the wavenumber of an ionised photoelectron scales as the square root of its kinetic energy, this has a limited contribution to EXAFS measurements.
In the XANES region, however, the source size can still have a considerable effect to the resolution. One way to mitigate the effect is to upgrade the source to a mini-focus X-ray tube. In addition, since the source size effect manifests as an energy gradient in the beam focus at the detector, an alternative solution would be to use a position-sensitive detector to compensate for the energy dispersion. Such a technique has been demonstrated to improve the energy resolution of bent \cite{honkanen14b} and diced analyser crystals \cite{huotari05,huotari06} where similar photon energy dispersions are inherent.

\begin{table}[]
\begin{tabular}{r|ccc}
\begin{tabular}[c]{@{}r@{}}Bragg angle\\(degrees)\end{tabular} & \begin{tabular}[c]{@{}c@{}}Source size\\($\times10^{-4}$)\end{tabular} & \begin{tabular}[c]{@{}c@{}}Other geometric\\($\times10^{-4}$)\end{tabular} & \begin{tabular}[c]{@{}c@{}}Total\\($\times10^{-4}$)\end{tabular} \\ \hline
85                                                              & 0.8                                                                     & 0.1                                                                         & 0.9                                                               \\
75                                                              & 2.6                                                                     & 0.8                                                                         & 2.9                                                               \\
65                                                              & 4.9                                                                     & 2.3                                                                         & 5.7                                                               \\
55                                                              & 8.2                                                                     & 5.4                                                                         & 10.4                                                             
\end{tabular}
\caption{Simulated geometric contributions to the relative energy resolution
($\Delta E/E$) for selected Bragg angles (full width at half maximum, FWHM). The diameter of the active area of the monochromator was assumed to be 60~mm. The intrinsic energy resolution of an analyser can be included by adding in quadrature to obtain the total energy resolution.  \label{tbl:rel_e_res}}
\end{table}

\subsubsection{Detectors and counting electronics}

The spectrometer is designed to be flexibly used with detectors for transmission, fluorescence, and imaging modes. Simultaneous data acquisition with multiple detectors is possible.
For the transmission mode measurements, typically either a NaI scintillator detector (2.5~cm diameter, active surface area of 500~mm$^2$) or a Amptek FastSDD (70~mm$^2$) can be used. The time constant of the scintillator setup is 2.8~$\upmu$s, allowing count rates up to $10^5$ counts/s, and the FastSDD can be used with $0.2$~$\upmu$s time constant for count rates up to $2\times10^6$ counts/s.
For the fluorescence mode measurements the FastSDD can be placed on the monochromator side of the sample.
In addition to the more traditional detectors, the instrument has a readily available imaging option utilising Advacam Modupix detector. The detector is build around the silicon Timepix chip \cite{llopart07} consisting of 256$\times$256 array of 55~$\times$~55~$\upmu$m$^2$ pixels. 

\subsubsection{Motors and Movement} 

\begin{figure}
\centering
\includegraphics[width=\columnwidth]{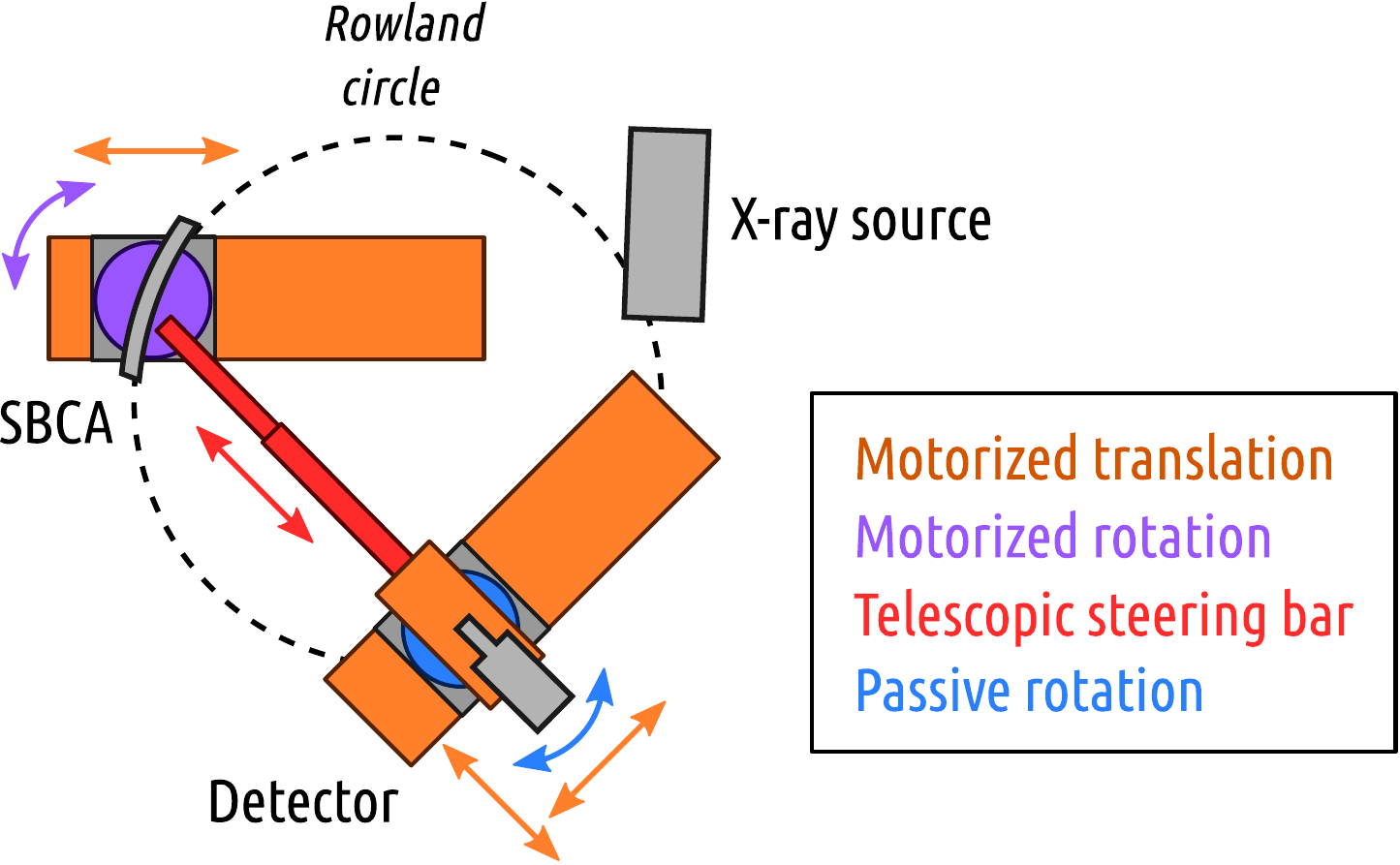}
\caption{The motor configuration and degrees of freedom. \label{fig:motors}}
\end{figure}

In the current setup, the x-ray source is kept fixed in position, and the monochromator and the detector follow the Rowland circle by three motorised linear translation stages, one motorised goniometer, and a passive rotation stage with a telescopic steering bar as presented in Figure~\ref{fig:motors}. 

The Bragg angle $\theta_B$ of the monochromator crystal is adjusted with a one-axis goniometer. 
The analyser-goniometer complex rests on a linear stage which adjusts the source-monochromator distance $\rho$ according to the relation $\rho = R_{b} \sin \theta_B$, where $R_{b}$ is the bending radius of the SBCA (or the diameter of the Rowland circle). In addition to the motorised degrees of freedom, the analyser stage has three manual actuators to adjust the tilt and the height of the analyser crystal, and the position of the crystal surface with respect to the rotation axis.

The distance of the detector from the monochromator is adjusted with a motorised linear stage which is installed on the freely rotating stage. A telescopic steering bar keeps the detector directed at the monochromator. The rotating stage is installed on another linear translation stage which follows the monochromatized beam.
Motors and data acquisition is run on SPEC version 6 by Certified Scientific Software.

\subsubsection{Sample Exchanger}

\begin{figure}
\centering
\includegraphics[width=0.75\columnwidth]{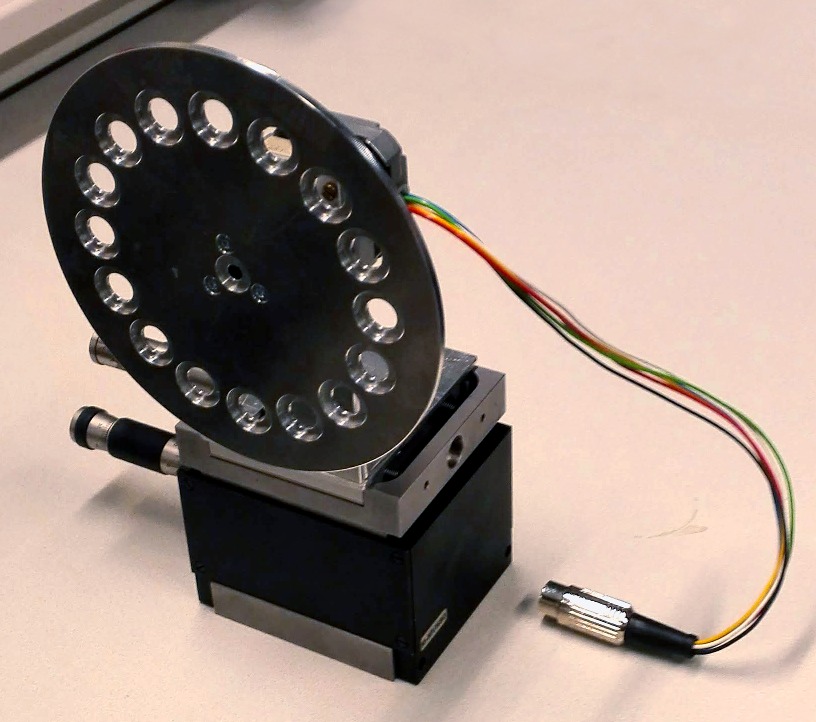}
\caption{The sample exchanger wheel. \label{fig:sample_changer}}
\end{figure}

The spectrometer is equipped with a sample exchanger which allows automated batch measurements. For transmission measurements, we use a stepper motor controlled circular plate design depicted in Figure~\ref{fig:sample_changer} positioned at the exit of the X-ray tube. 
The sample exchanger can also be positioned in front of the detector, in the monochromatic beam, for radiation sensitive samples and for fluorescence measurements.

\subsubsection{Enclosure and He Chamber} 

\begin{figure}
\centering
\includegraphics[width=\columnwidth]{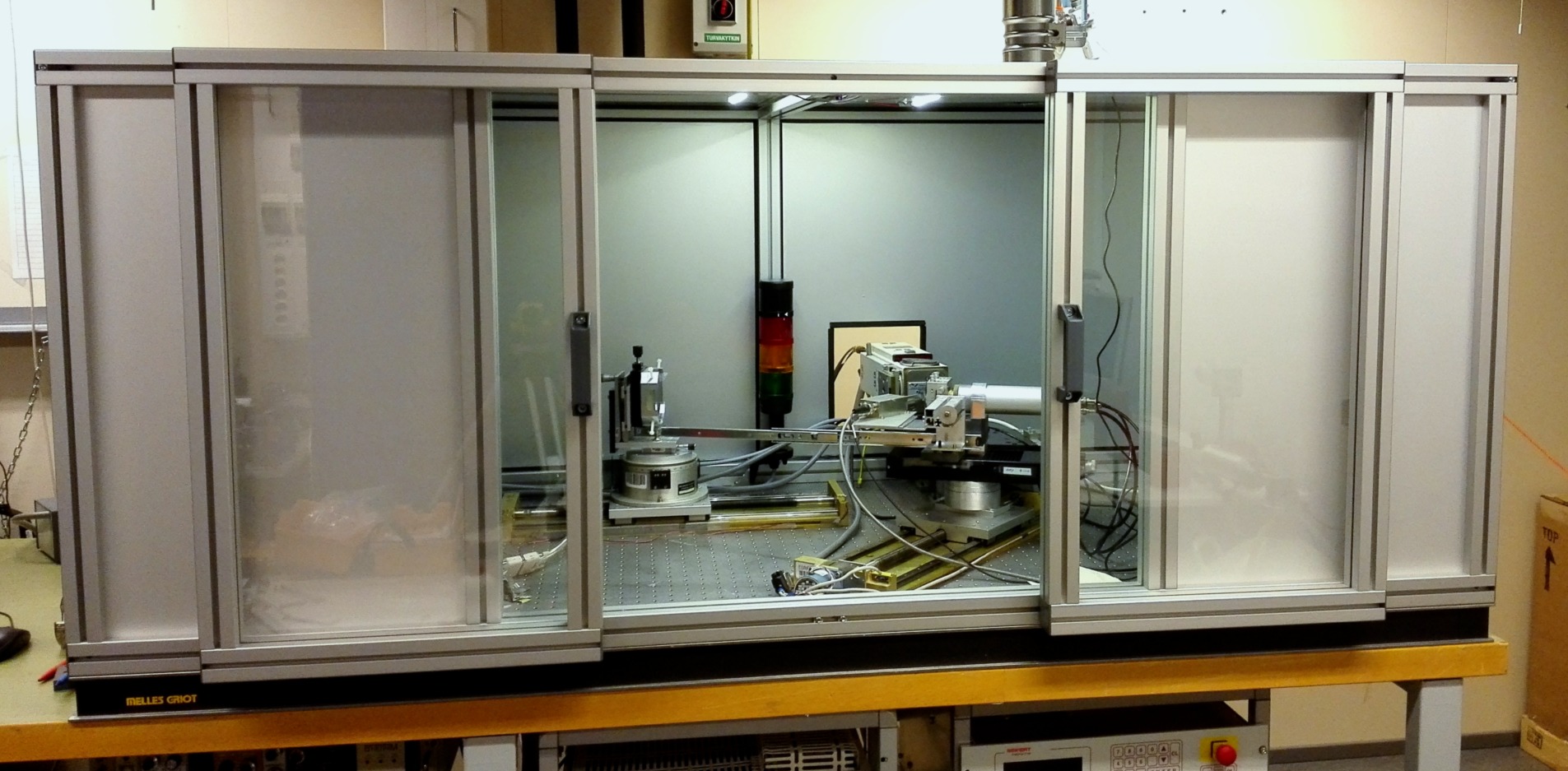}
\caption{The instrument enclosure.\label{fig:enclosure}}
\end{figure}

The instrument is enclosed in a cabinet pictured in Figure~\ref{fig:enclosure}.  
User safety is ensured by specifically positioned lead shielding to absorb the directed beams, whereas the aluminium walls of the enclosure are itself thick enough to block any stray scattering and fluorescent radiation. The windows of the cabinet doors are lead glass. The dimensions of the enclosure are $200 \times 100 \times 80$~cm$^{3}$.

\begin{figure}
\centering
\includegraphics[width=\columnwidth]{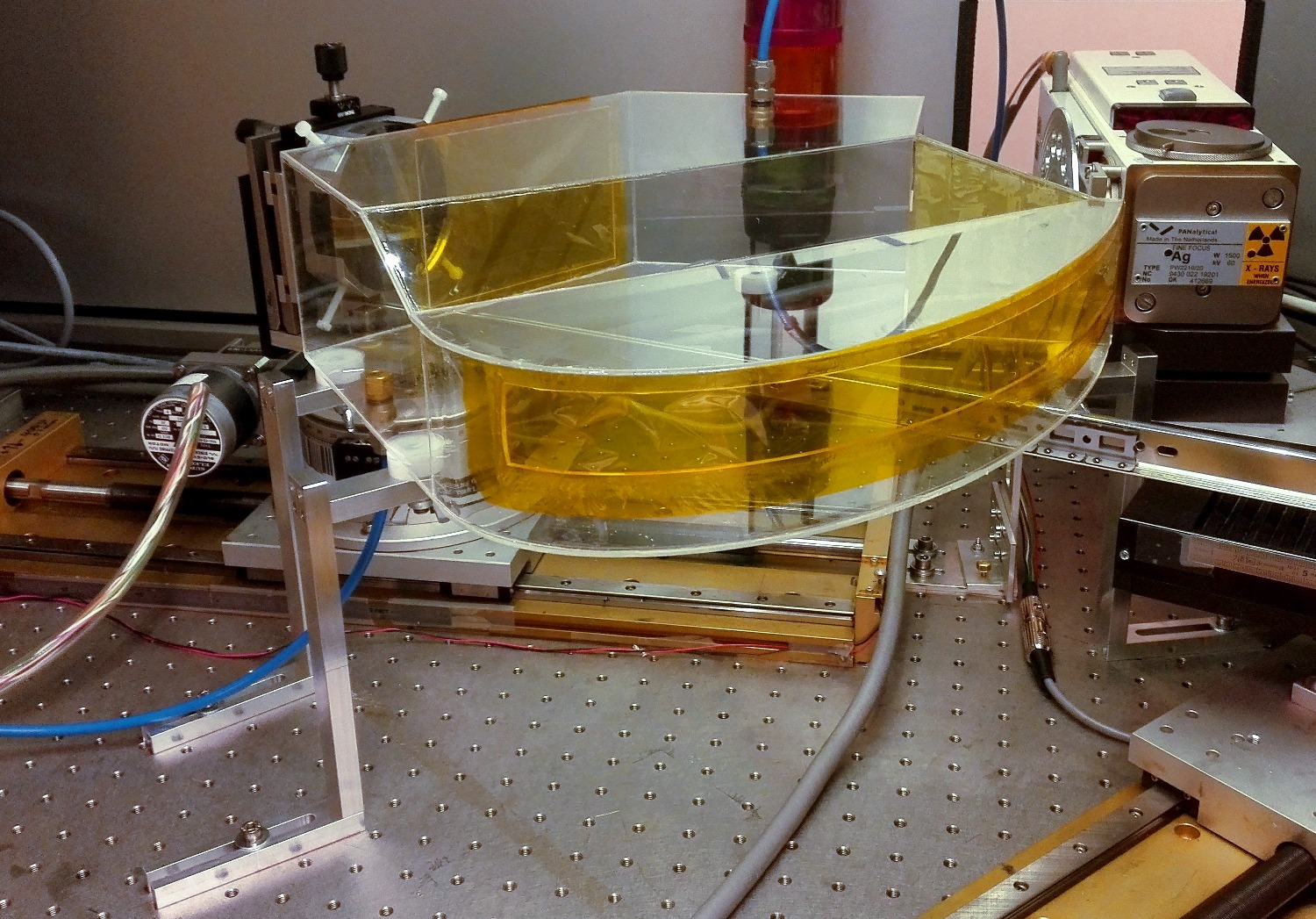}
\caption{The helium chamber.\label{fig:heliumbox}}
\end{figure}

In order to reduce the scattering and absorption by the air, the instrument is fitted with a Kapton-windowed helium chamber which is presented in Fig.~\ref{fig:heliumbox}.

\subsection{Operation}

The purpose of an XAS instrument is the determination of the absorption coefficient $\mu x$ as the function of photon energy either directly from the attenuation of the beam transmitted through the sample or indirectly from the fluorescence yield.

\begin{figure}
\centering
\includegraphics[width=0.85\columnwidth]{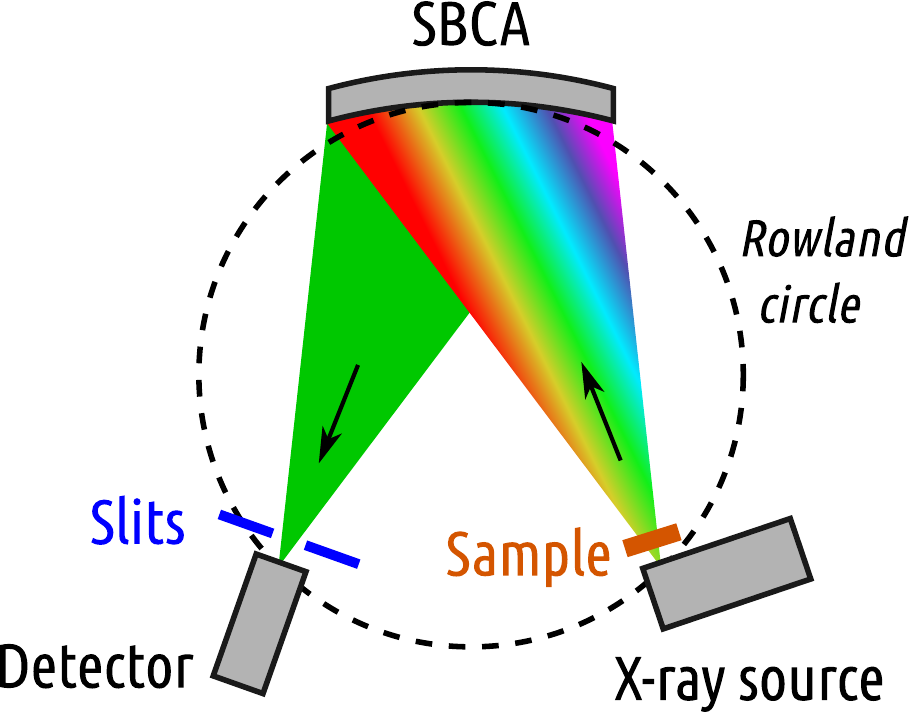}
\caption{Direct absorption measurement in transmission mode. Alternatively, the sample can be placed in front of the slits.\label{fig:transmission}}
\end{figure}

The transmission configuration is presented in Figure~\ref{fig:transmission}. 
For transmission measurements the sample can be positioned either at the exit of the X-ray tube or in the front of the detector. The advantage with the former is the stability of the beam footprint; owing to the spherical aberration, the vertical and horizontal focal lengths differ considerably at the detector side as a function of the Bragg angle. Therefore the footprint of the beam through the sample varies during the scan which might cause artefacts to the spectrum if the optical thickness of the sample is not uniform. At the exit of the X-ray tube the beam footprint is stable but since the beam is polychromatic, the dose rate is considerably larger which may harm radiation sensitive samples.

Since modern X-ray tubes and their power sources are highly stable, the measurements of the direct and transmitted beams can be conducted separately using the same detector without the need for an additional beam monitor. Furthermore, since $\mu x$ depends on the beam intensities in a highly non-linear manner, the background levels owing to the air scattering and the fluorescence of the components need to be determined and subtracted from the signals proper. This is done by offsetting the detector from the beam focus and repeating the scan. To obtain an accurate estimate of the background at the focal spot, the offset measurement is performed on both sides of the focus.
A low-order polynomial is fitted to the mean of the background signals and the fit is subtracted from the recorded signals. The absorption coefficient is thus obtained from the Beer-Lambert law by
\begin{equation}
\mu x = - \ln \frac{I - y_{bg}}{I_0 - y_{0,bg}}, 
\end{equation}
where $I_0$ and $I$ are the direct and the transmitted beam intensities and $y_{0,bg}$ and $y_{bg}$ are the polynomial fits to their background signals, respectively.

\begin{figure}
\centering
\includegraphics[width=0.85\columnwidth]{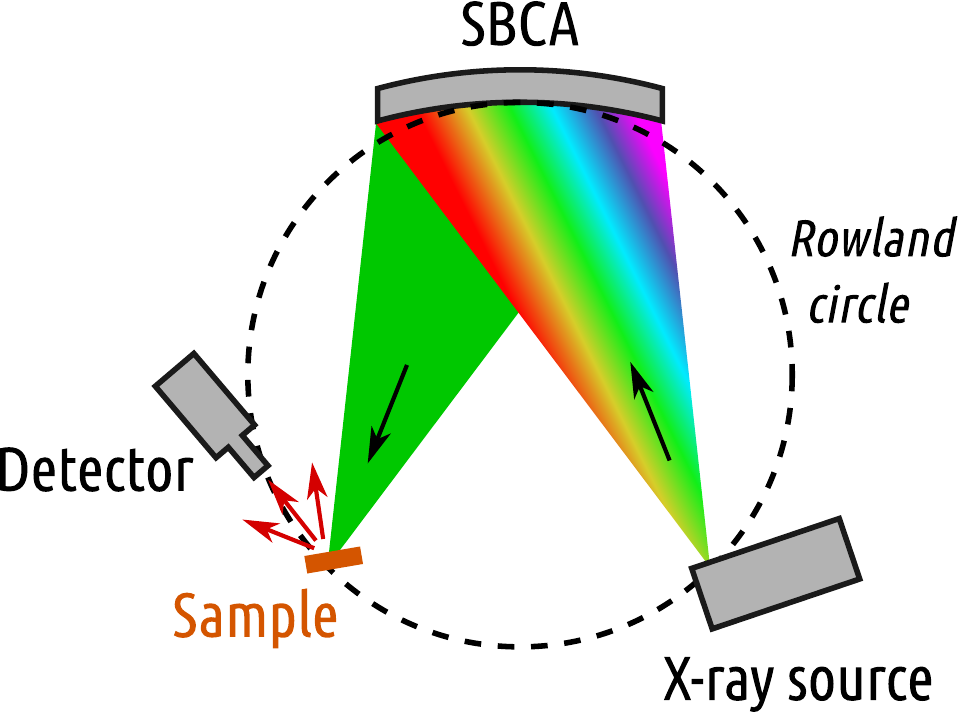}
\caption{Indirect absorption measurement in the fluorescence mode.\label{fig:fluorescence}}
\end{figure}

The fluorescence mode configuration is presented in Figure \ref{fig:fluorescence}. As with the transmission measurements, the signal from the sample and the reference signal proportional to the intensity of the incident beam are usually measured separately. As the reference signal for the element with the atomic number $Z$, one may use fluorescence from a $Z-1$ foil. Alternatively, if the absorption of the sample and the possible substrate is low, one could measure the reference signal from the transmitted beam with the scintillator. However, since the detector and thus the beam spot moves on the sample, this may introduce geometrical contributions to the spectrum which do not normalise out properly.

For optically thin sample ($\mu x \ll 1$), the absorption coefficient is directly proportional to the fluorescence signal:
\begin{equation}
\mu x \propto I_{f}/I_{ref}
\end{equation}
where $I_{f}$ is the fluorescence signal from the sample and $I_{ref}$ is the reference signal. 
The fluorescence mode can be used also for thick samples but in that case the data has to be corrected for self-absorption.

The Python implementations of the data extraction routines are available at \url{https://github.com/aripekka/helxas}.

\section{Use cases}

\subsection{XANES}

The most standard application of the instrument is to measure X-ray absorption near-edge structure (XANES) in the transmission mode. As a benchmark, we measured the K edge absorption spectrum of 10~$\upmu$m thick cobalt foil using Si(533) monochromator. The X-ray tube acceleration voltage was 20~kV and the current 2~mA. 
Fairly low tube power was used in order to avoid the saturation of the scintillator detector with the direct beam.
The average direct beam count rate normalised to the tube current was 20~kcps/mA 
of which the background contribution was 0.6--1~kcps/mA,  
the rate decreasing towards the higher photon energies.  
The count rate of the beam transmitted through the foil was 13~kcps/mA below the Co K edge and 
0.9~kcps/mA above. 
The background contributions were 0.2--0.3~kcps/mA,  
the rates decreasing towards the higher photon energies. The intensity of the beam without sample was measured 2 times by repeated scans and 2 times with the sample, each scan consisting of 301 energy points with a counting time of 5 seconds per point. The background measurements were done in the same manner for the transmitted and the direct beam, totalling 4 scans. Total counting time not including motor movement was thus 3.3 hours. The relative statistical uncertainty in the obtained $\mu x$ is less than 1.0\%. We note that by replacing the scintillator with a high count rate detector like a silicon drift diode, which would not saturate even for the maximum tube current of 40 mA, we could obtain 0.3\% statistical accuracy in less than 1.5 hours.

\begin{figure}
\centering
\includegraphics[width=\columnwidth]{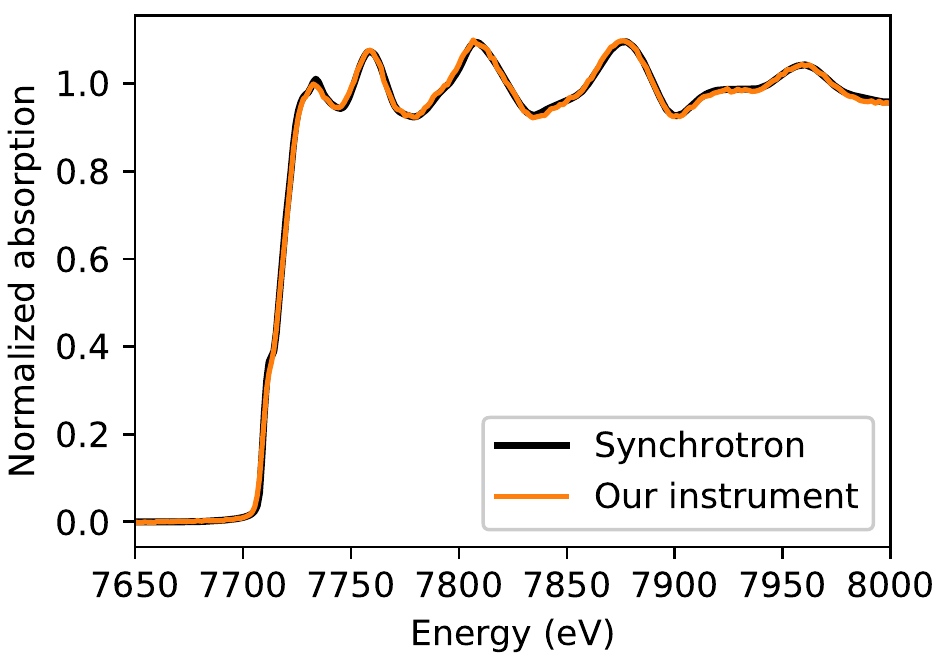}
\caption{Normalised K edge absorption spectrum of a cobalt foil compared with a synchrotron reference measured at NSLS X11B \cite{ravel05}.\label{fig:Co_K_edge_XANES}}
\end{figure}

The result is presented together with the literature data obtained at a synchrotron \cite{ravel05} in Figure~\ref{fig:Co_K_edge_XANES}. The measured spectrum corresponds well to the reference over the 300 eV range. Slight deviations, especially at the main edge, are most likely due to the energy resolution of our instrument, estimated to be of the order 2.5~eV at the absorption edge, is not quite as high as the one of a synchrotron beam line but more than sufficient for most applications.

At the time of writing, the highest edge in energy measured and published in transmission mode is the U L3 edge at 17.2 keV using Ge(999) monochromator \cite{bes18}. The lowest energy data that is published is taken at Mn 6.5 keV using Si(440) \cite{wang17,kuai18} and the lowest measured is Sb L1 edge at 4.7 keV using Si(400). 

\subsection{EXAFS: long $k$-range measurement}

In the normal configuration the lowest attainable Bragg angle is around 67$^\circ$ which is limited by the range of the larger translation stage of the detector (see Fig.~\ref{fig:motors}). However, the range can be considerably extended by performing the measurement in parts for different translation stage positions. This makes it possible to measure extended X-ray absorption fine-structure (EXAFS) up to 1000 eV or more above the absorption edge.  

The EXAFS acquisition was demonstrated with a 5~$\upmu$m thick nickel foil using a Si(551) monochromator. The spectrum was measured over two Bragg angle ranges: 81--67$^\circ$ and 69--57$^\circ$. The overlap of the ranges was to assure the proper stitching of the parts. The X-ray tube voltage and the current for the ranges were 10~kV/5~mA and 10~kV/10~mA, respectively. Unusually low acceleration voltage was used to avoid the excitation of W L$_3$ electrons at 10.2~keV and thus the presence of W L$\alpha_{1,2}$ lines which overlap with the Ni K edge. The presence of tungsten in the tube is most likely owing to the evaporation of the filament. 

The current-normalised count rates of the direct beam were 0.6--4~kcps/mA and the background rates 0.1--0.3~kcps/mA, both decreasing toward the higher photon energies. The count rates of the transmitted beam were 3~kcps/mA below the Ni K edge and 0.3--0.9~kcps/mA above. The background rates were 0.05--0.2~kcps/mA. The direct and the transmitted beams in the low-energy part were measured 15 and 30 times, respectively, with 401 points per scan and 5 seconds counting time per point. In the high-energy part the numbers of scans were 10 and 21, for the direct and the transmitted beams respectively, with 301 points and 10 seconds per point. The total counting time including the background scans was 53.2 hours.  

\begin{figure}
\centering
\includegraphics[width=\columnwidth]{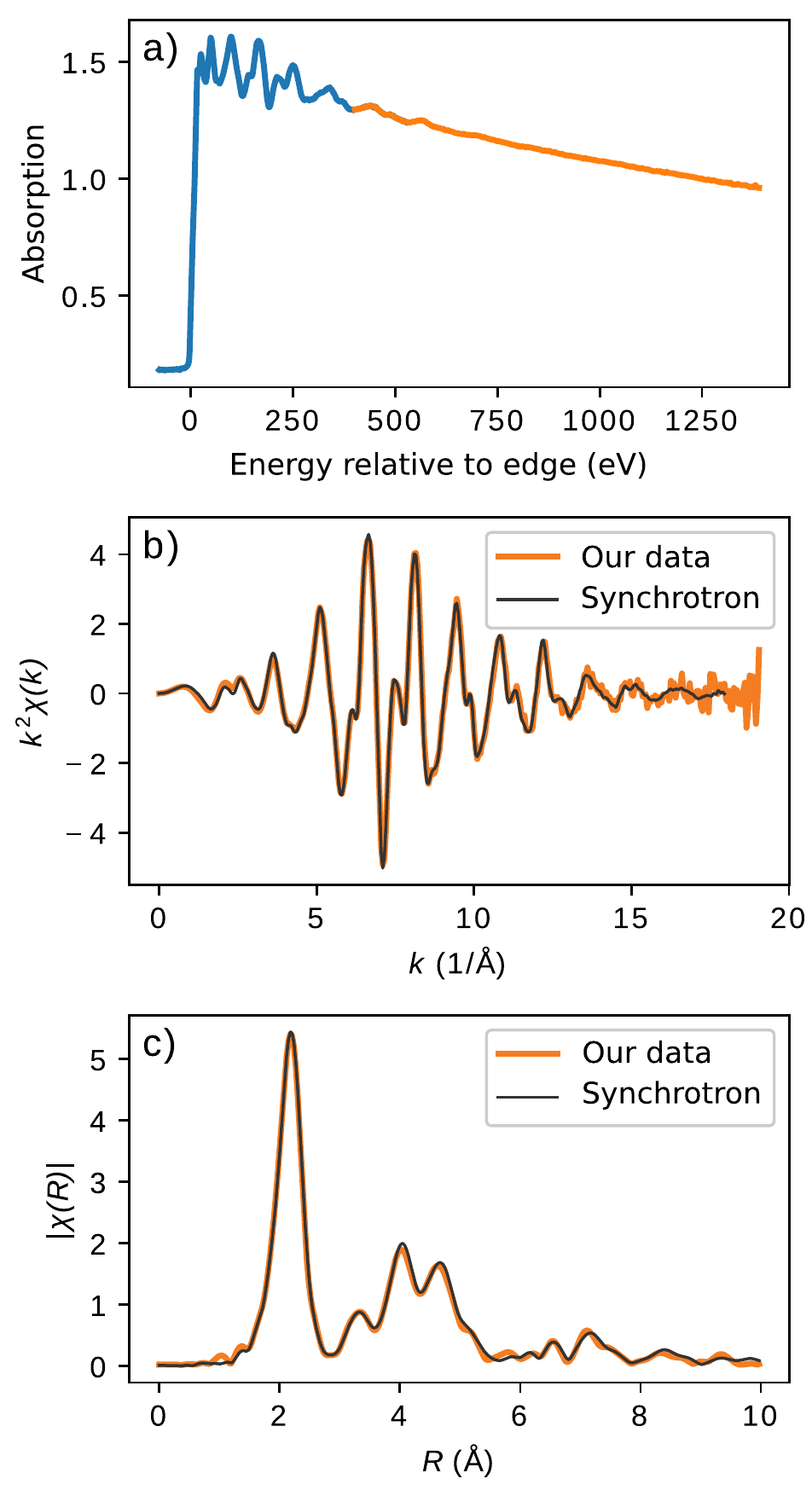}
\caption{The Ni K edge EXAFS spectra of a Ni metal foil.
 \emph{a)} The raw absorption spectrum measured in two parts. \emph{b)} The extracted EXAFS oscillations. \emph{c)} The Fourier transform of the EXAFS oscillations. The synchrotron reference was measured at APS 13ID \cite{ravel05}.\label{fig:EXAFS}}
\end{figure}

The measured absorption spectrum is presented in Figure~\ref{fig:EXAFS} a. With the two-part measurement, the EXAFS region can be obtained up to 1400 eV above the edge, which corresponds almost 20 1/\AA \ in $k$-space, as seen in Figure~\ref{fig:EXAFS} b. The EXAFS signal and its Fourier transform compare well to the synchrotron literature reference \cite{ravel05} in quality (Figures~\ref{fig:EXAFS} b and c).

\subsection{Detection in fluorescence mode}

The fluorescence mode was tested with two kinds of samples: a 5~$\upmu$m thick Co metal foil and a 100~nm thick thin films of CoO and Co$_3$O$_4$. The fluorescence yield was measured with an Amptek XR-100CR with an active area of 25 mm$^2$. The detector was placed at about 1.5~cm distance from the sample, corresponding to 0.1~sr in the covered solid angle. 

The spectrum of the Co foil was scanned 8 times, 201 points per scan. The counting time was 10 s/point, totalling 4.5 hours. The total intensity of Co K$\alpha$ fluorescence signal above the edge was 1100 counts/s of which the background contribution was 330 counts/s. In order to normalise the signal, the fluorescence of an iron foil was measured over the same energy grid. The Fe foil was scanned one time, 10 s/point, totalling 0.6 hours. The count rate was 1000 photons/second. The tube voltage was 20~kV and the current was 40~mA in both cases. 

\begin{figure}
\centering
\includegraphics[width=\columnwidth]{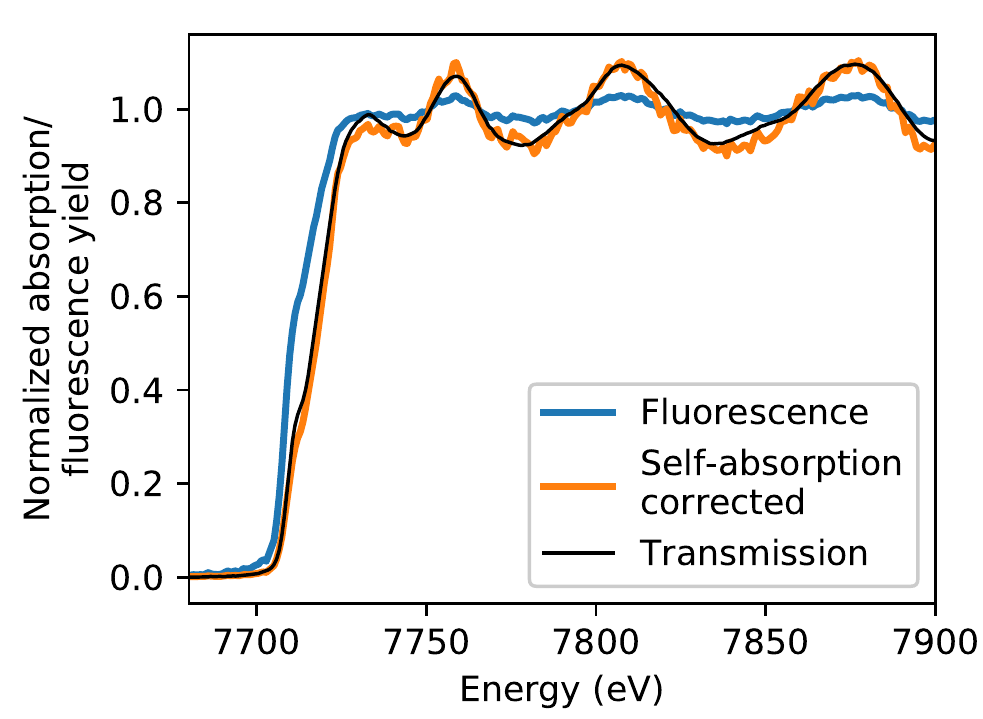}
\caption{Cobalt K$\alpha$ fluorescence yield spectrum measured in fluorescence mode from Co foil (blue), the same signal after self-absorption correction (orange) and the absorption spectrum measured in transmission (black).
\label{fig:Co_K_fluorescence}}
\end{figure}

For the normalisation, the Fe signal was smoothed with a polynomial fit. The measured fluorescence was corrected for the self-absorption using Athena \cite{ravel05}. For correction, the incidence angle value was set to 90$^\circ$ and outgoing value was chosen to be 25$^\circ$. The measured and corrected fluorescence signals are presented together with Co absorption spectrum measured in transmission mode in Figure~\ref{fig:Co_K_fluorescence}. The self-absorption corrected fluorescence signal is found to follow the transmission spectrum as expected.

In addition to the Co foil, we measured 100~nm thick CoO and Co$_3$O$_4$ films \cite{kim17} grown on a silicon substrate using atomic layer deposition \cite{suntola89,leskela02}. 
The CoO film was scanned 6 times, 201 points per scan. The counting time per point was 50 seconds, totalling 16.8 hours of counting time. The count rate above the edge was 25 photons/second, of which the background contribution was
15 photons/s. The Co$_3$O$_4$ film was scanned in the same manner 7 times, with  19.5 hours of total counting time. The count rate above the edge was 30 photons/second, of which the background contribution was
15 photons/s. The tube voltage and current were 20~kV and 40~mA, respectively.

\begin{figure}
\centering
\includegraphics[width=\columnwidth]{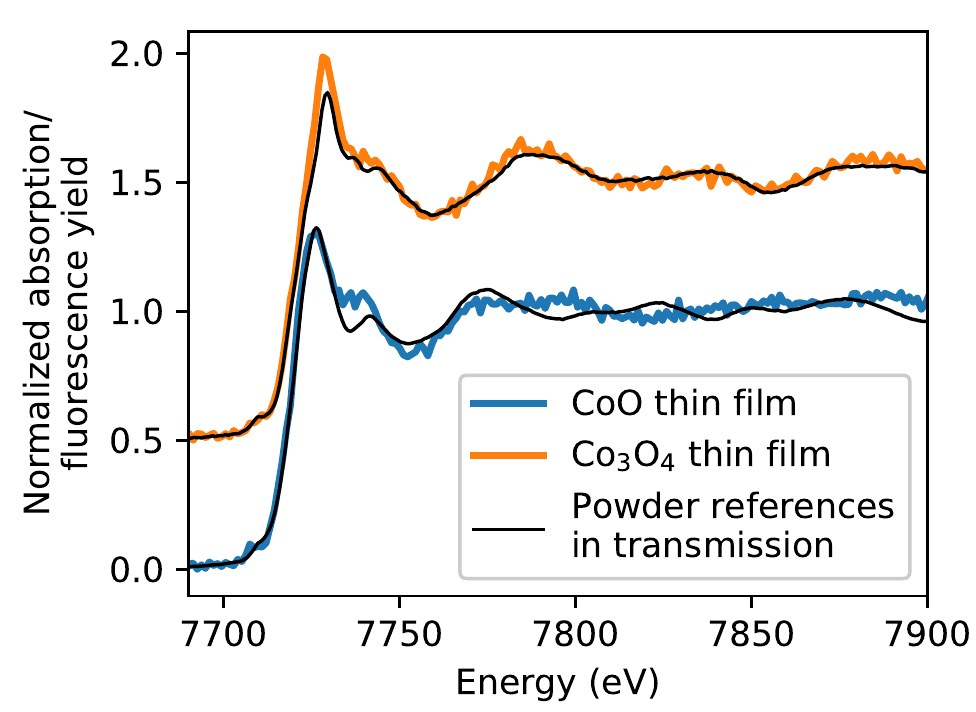}
\caption{Cobalt K edge spectrum measured in fluorescence mode from 100 nm 
thick CoO (blue) and Co$_3$O$_4$ (orange) thin film grown with atomic layer deposition. The reference powders spectra measured in transmission mode are shown with black lines. The Co$_3$O$_4$ spectra are shifted along the $y$-axis for clarity. \label{fig:thin_film_spectrum}}
\end{figure}

The thin film spectra together with the powder references measured in transmission are presented in Fig.~\ref{fig:thin_film_spectrum}. In the case of Co$_3$O$_4$ film, the data follows the reference spectra closely. CoO deviates slightly from the reference which might be due atomic level differences between the thin film oxide and the reference, or due to the post-preparation oxidation of the film.

Assuming that the densities of CoO and Co$_3$O$_4$ in the films were 6.44 and 6.11~g/cm$^3$, we can estimate the edge steps $\Delta \mu x$ to be 0.016 and 0.014, respectively. In the setup used, this corresponds to the normalized count rate 600--1100 photons/s for unit edge step if the self-absorption is neglected. The estimate is in accordance with the measured signal strength of 770 photons/s from the Co foil ($\Delta \mu x = 1.4$). 

\begin{figure}
\centering
\includegraphics[width=\columnwidth]{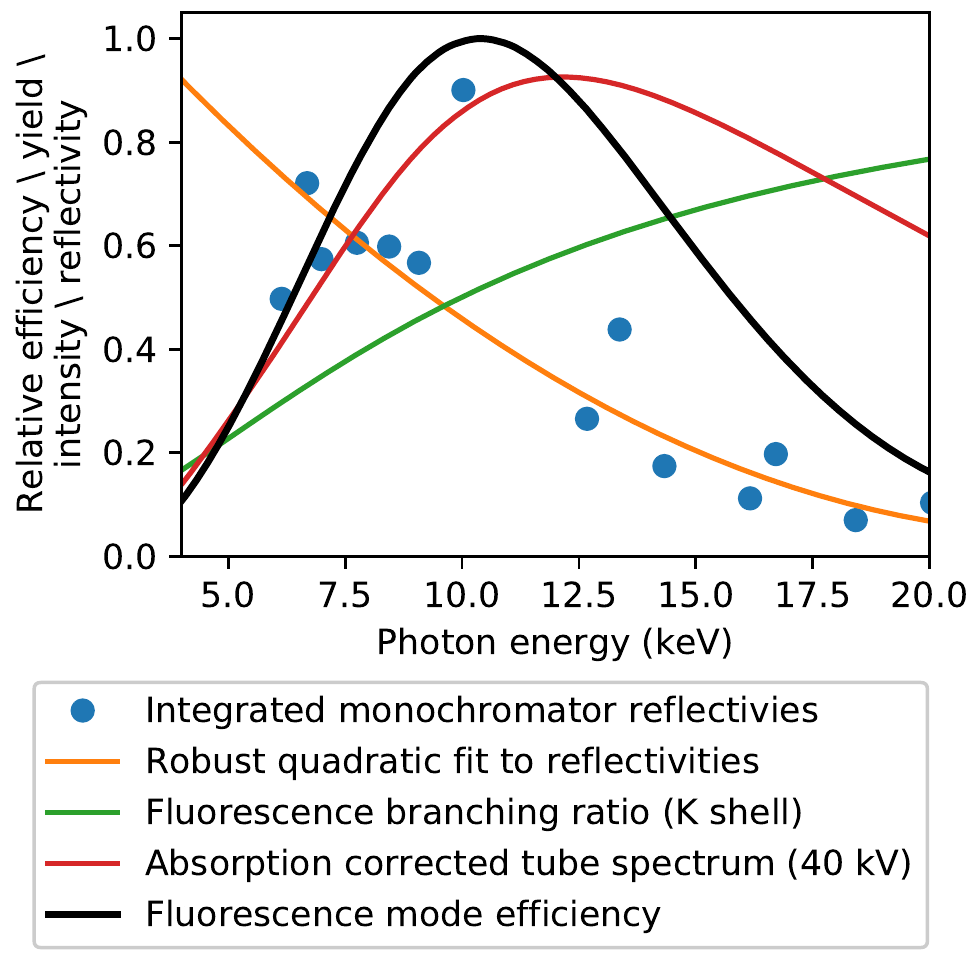}
\caption{Estimated efficiency of the fluorescence mode as a function of K edge energy. The efficiency is a product of the tube spectrum, monochromator reflectivity and the branching ratio of radiative and non-radiative core hole decay. The background noise is not included in the estimate.  \label{fig:fluorescence_mode_efficiency}}
\end{figure}

To extend the signal strength per unit edge step to other K edge energies, we have estimated the expected fluorescence yield per unit edge step (without background noise) in Figure~\ref{fig:fluorescence_mode_efficiency}. The estimate is composed of three components. The spectrum of the X-ray tube was simulated according to \cite{ebel99} and corrected for absorption by the Kapton windows and helium in the helium chamber and by the air surrounding the chamber. The acceleration voltage of 40 kV was used in the calculations as that would be used for heavier elements. With a fixed tube current, the intensity at Co K is approximately same for the simulated 40 kV spectrum and 20 kV used in the measurements, so the measured count rate for unit edge step applies. The integrated monochromator reflectivities were calculated for a selected reflections of 0.5 m spherically bent silicon with wafer thickness of 180 $\upmu$m at 75$^{\circ}$ Bragg angle using 1D Takagi-Taupin solver \emph{pyTTE} (\url{https://github.com/aripekka/pyTTE}) and the quadratic trend was obtained with robust fitting. The branching ratios for radiative decay of core hole states were obtained using \emph{xraylib} \cite{schoonjans11}. The product of these three factors gives the relative efficiency of the fluorescence mode for K shells per unit edge step of absorbing material which can be used to estimate the expected signal strength when absorption of the material is known and the self-absorption is negligible. In terms of fluorescence, the optimal energy for our instrument is around 10~keV with full width at half maximum of 8~keV. However, it should be noted that the different reflections may deviate considerably from the general trend in terms of their reflectivity and that especially absorption of lower energy photons is very sensitive to changes in the measurement environment. In addition, the quantum efficiency of the fluorescence detector needs to be taken into account.

To our knowledge this is the first time when X-ray absorption spectrum is measured \emph{via} fluorescence using a laboratory-scale instrument. 

\subsection{Imaging}

The combination of a monochromatic beam and a position-sensitive detector allows us to utilise absorption edge subtraction as a contrast mechanism in transmission imaging thus allowing element-sensitive mapping of the sample. The idea of absorption edge subtraction imaging is to take absorption images slightly above and below the absorption edge of interest. The sharp increase in the absorption gives a strong contrast to the element in question in the subtraction of the above and below edge images \cite{lewis97}. 

\begin{figure*}
\centering
\begin{subfigure}[b]{0.221\textwidth}
\includegraphics[width=\textwidth]{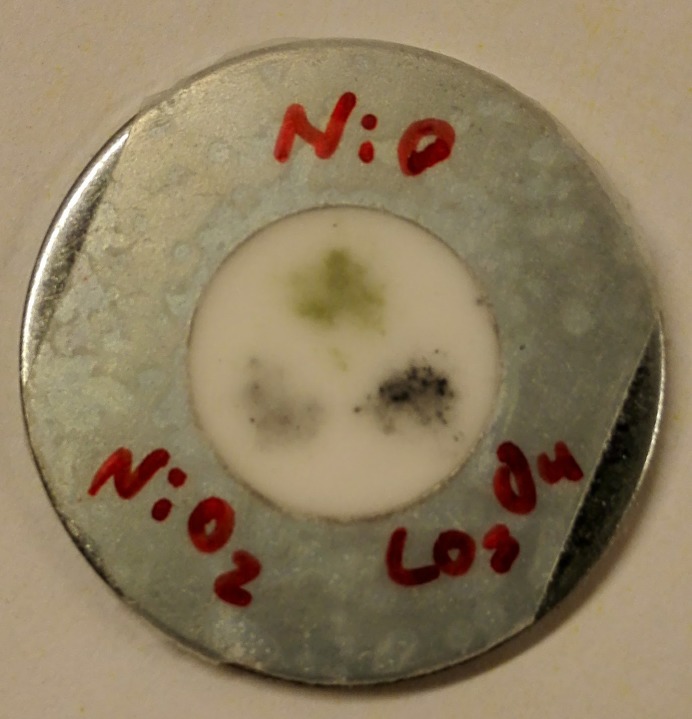}
\caption{Sample\label{fig:imaging_sample}}
\end{subfigure}
~
\begin{subfigure}[b]{0.23\textwidth}
\includegraphics[width=\textwidth]{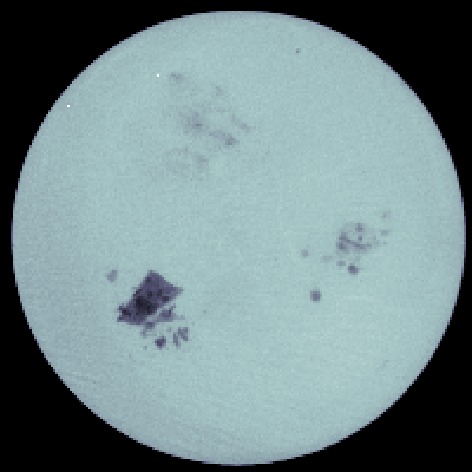}
\caption{Below Co K edge\label{fig:imaging_below_K}}
\end{subfigure}
~
\begin{subfigure}[b]{0.23\textwidth}
\includegraphics[width=\textwidth]{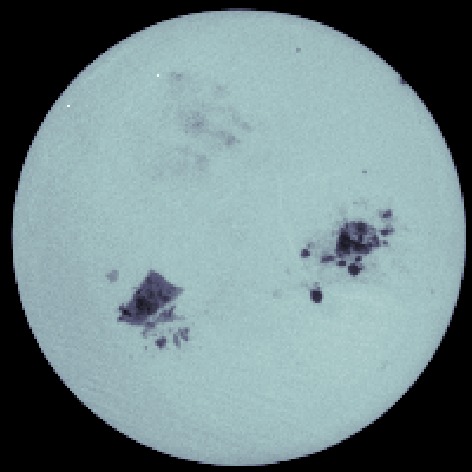}
\caption{Above Co K edge\label{fig:imaging_above_K}}
\end{subfigure}
~
\begin{subfigure}[b]{0.23\textwidth}
\includegraphics[width=\textwidth]{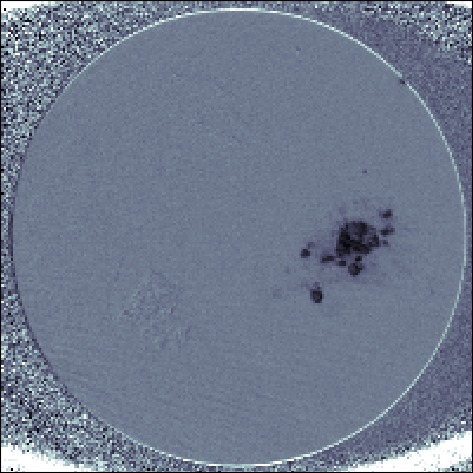}
\caption{Difference\label{fig:imaging_difference}}
\end{subfigure}
\caption{Absorption edge subtraction imaging using a position-sensitive detector. The images of the sample (powders of NiO, NiO$_2$ and Co$_3$O$_4$ embedded in potato starch) were taken at photon energies above and below the absorption edge allowing element-sensitive imaging.}
\end{figure*}

The test sample was a steel washer filled with potato starch embedded with small amounts of NiO, NiO$_2$, and Co$_3$O$_4$ powders (Fig.~\ref{fig:imaging_sample}). The inner diameter of the washer was 11~mm and thickness was 1~mm. The sample was attached in front of the detector and the detector was placed into the cone of the monochromatic beam so that the detector chip was fully illuminated. Two images were taken of the sample, one below and another above the Co K-edge. Another two images were taken without the sample (\emph{i.e.} the flat field images) at the corresponding photon energies. Due to the low noise of the detector, the dark field images without the beam could be omitted. The transmission images obtained by taking the logarithm of the flat field normalised frames are presented in Figures~\ref{fig:imaging_below_K} and \ref{fig:imaging_above_K}. The distribution of cobalt is then extracted by subtraction of the below and above edge images. The result is presented in Fig.~\eqref{fig:imaging_difference}.

This proof-of-concept study indicates that element-sensitive imaging is a viability with a Johann type laboratory spectrometer. With further development the method can be potentially used to study not only the distribution of elements in the sample but also their chemical state with proper XAS analysis of the data of each pixel.
The presented method can provide valuable information on a laboratory scale in, for example, \emph{in situ} battery and \emph{in vivo} biological studies. In addition, the absorption edge subtraction imaging can be readily combined with computed tomography to obtain 3D reconstruction with element contrast of the sample.

\section{Applications}

In this section we present various real-life projects for which our instrument has been used. The purpose of the showcase is to highlight the advantages of having a laboratory scale instrument readily available.

The development and synthesis of new electrochemical materials for energy devices is an active and rapidly advancing field of materials science. X-ray absorption spectroscopy would be a valuable tool in understanding the chemical composition and oxidation states of newly synthesised materials but it suffers from a scarcity of synchrotron beamtime available for materials characterisation as well as a lengthy and uncertain beamtime application process. Such standard measurements are an ideal use for our instrument which we demonstrated recently in cases of mesoporous MnCo$_2$O$_4$ and LaMnO$_{3+\delta}$ electrocatalysts \cite{wang17,kuai18}, where we used K edge absorption spectroscopy to determine the oxidation states of Mn and Co in the bulk of the prepared nanoparticles.

\begin{figure}
\centering
\includegraphics[width=0.95\columnwidth]{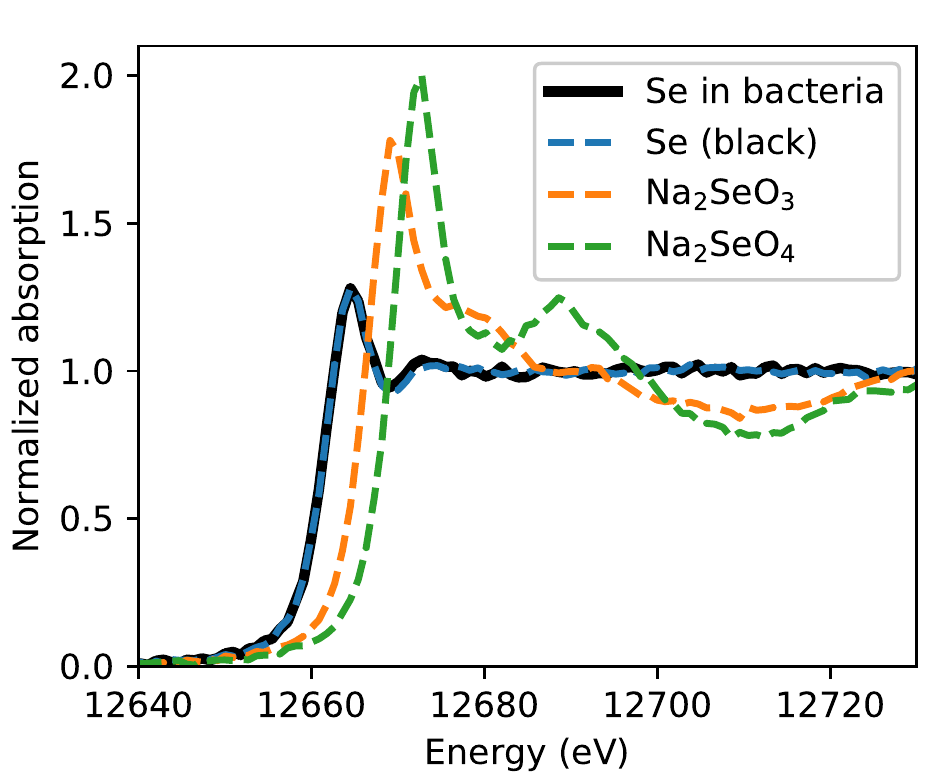}
\caption{Se K edge of accumulated selenium in \emph{Pseudomonas} bacteria compared with reference spectra. The acceleration voltage and current of the X-ray tube were 20~kV and 10~mA, respectively. The total counting time for the presented dataset was 39 hours. \label{fig:bacteria}}
\end{figure}

X-ray absorption spectroscopy is also suitable for non-invasive biochemical studies such as investigating the binding and activity of metallic ions in proteins and peptides \cite{lima14,alies16} or metabolism of toxic elements in plants and animals \cite{lombi09, misra10}. With our instrument we have investigated the bioaccumulation of selenium in \emph{Pseudomonas sp.} strain T5-6-I previously isolated from an ombrotrophic boreal bog \cite{lusa16}. The bacteria cultivated in SeO$_{3}^{-2}$ solution were freeze-dried and Se K edge was measured from them in the dormant state. The preliminary results presented in Figure~\ref{fig:bacteria} show that the selenium is accumulated by \emph{Pseudomonas sp.} strain T5-6-I as Se(0) which is consistent with earlier studies \cite{lusa17}. The in-depth analysis is under preparation \cite{lusa_inprep}.

Actinides are a very important group of elements with regards to nuclear energy production and waste management.
XAS is widely used method for studying their chemistry and composition but its usage is severely limited by the lack of suitable beamlines. It was recently demonstrated that laboratory scale spectrometers are a viable option for actinide research by measuring uranium L$_3$ edges from several oxides of uranium \cite{bes18}. Due to their compact size, similar instruments could be integrated with the handling equipment of radioactive materials which allows the XAS studies of highly active materials in a safe manner.

A unique application of the laboratory scale XAS is the possibility to run extremely long \emph{in situ} chemical reaction studies. We have studied \cite{moya-cancino19} the chemistry of Co nanoparticles used to catalyse the formation of long hydrocarbon chains from hydrogen and carbon monoxide in Fischer-Tropsch synthesis. The longest \emph{in situ} runs have lasted over 200 hours which excludes the possibility of conducting the same experiment at synchrotrons. In such experiments laboratory based instruments are complementary to synchrotron studies.

\section{Conclusions}

In this work we have introduced our Johann type X-ray absorption spectrometer which is a versatile and low-cost setup capable of measuring absorption spectra of most \emph{3d} K and \emph{5d} L edges in both transmission and fluorescence modes. In addition, the use of position-sensitive detector enhances the instrument with imaging capabilities. The presented real-life applications show that such instruments have a huge potential in addressing the need for synchrotrons and free electron lasers. In applications requiring long measurement times or potentially hazardous samples laboratory-scale instruments can truly complement the large-scale facilities.

\subsection*{Acknowledgements}

The authors want to thank Gerald T. Seidler for insightful discussions regarding the instrument design, Tomi Iivonen for providing the cobalt oxide thin film sample and Papa Sene for the CAD drawings of the instrument. The work was supported by the Academy of Finland (grant 1295696) and University of Helsinki Doctoral Program in Materials Research and Nanosciences (MATRENA). Funding to A.-J.K. was granted by the Finnish Cultural Foundation through the Shared Waters project.

\bibliography{refs}

\end{document}